\documentclass[reprint,amsmath,amssymb,amsart,aps,superscriptaddress,twocolumn,float]{revtex4-2}
\usepackage{graphicx,graphics}
\usepackage{dcolumn}
\usepackage{bm}
\usepackage{amssymb}
\usepackage{amsfonts}
\usepackage{float}
\usepackage{xcolor}

\begin{document}
	
\title{Helicity-tunable spin Hall and spin Nernst effects in unconventional chiral fermion semimetals XY (X=Co, Rh; Y=Si, Ge)} 

\author{Ting-Yun Hsieh}
\thanks{These authors contributed equally to this work.}
\affiliation{Department of Physics, National Taiwan University, Taipei 10617, Taiwan, Republic of China\looseness=-1}

\author{Babu Baijnath Prasad}
\thanks{These authors contributed equally to this work.}
\affiliation{Department of Physics, National Taiwan University, Taipei 10617, Taiwan, Republic of China\looseness=-1}
\affiliation{Nano Science and Technology Program, Taiwan International Graduate Program, Academia Sinica, Taipei 11529, Taiwan, Republic of China\looseness=-1}
\affiliation{Physics Division, National Center for Theoretical Sciences, Taipei 10617, Taiwan, Republic of China\looseness=-1}

\author{Guang-Yu Guo}
\email{gyguo@phys.ntu.edu.tw}
\affiliation{Department of Physics, National Taiwan University, Taipei 10617, Taiwan, Republic of China\looseness=-1}
\affiliation{Physics Division, National Center for Theoretical Sciences, Taipei 10617, Taiwan, Republic of China\looseness=-1}

\date{\today}

\begin{abstract}

Transition metal monosilicides CoSi, CoGe, RhSi and RhGe in the chiral cubic B20 structure (the CoSi family) 
have recently been found to host unconventional chiral fermions beyond spin-1/2 Weyl fermions, 
and also to exhibit exotic physical phenomena such as long Fermi
arc surface states, gyrotropic magnetic effect and quantized circular photogalvanic effect.
Thus, exploring novel spin-related transports in these unconventional chiral fermion semimetals 
may open a new door for spintronics and spin caloritronics. 
In this paper, we study the intrinsic spin Hall effect (SHE) and spin Nernst effect (SNE) in the CoSi family 
based on {\it ab initio} relativistic band structure calculations. 
First, we find that unlike nonchiral cubic metals, the CoSi family have two independent nonzero spin Hall (Nernst) conductivity
tensor elements, namely, $\sigma_{xy}^z$ and $\sigma_{xz}^y$ ($\alpha_{xy}^z$ and $\alpha_{xz}^y$) instead of one element.
Furthermore, the SHC ($\sigma_{xy}^{z}$ and $\sigma_{xz}^{y}$) and helicity of the chiral structure are found to be correlated, thus enabling SHE detection of structural helicity and also chiral fermion chirality.
Second, the intrinsic SHE and SNE in some of the CoSi family are large.
In particular, the calculated spin Hall conductivity (SHC) of RhGe is as large as -140 ($\hbar$/e)(S/cm).
The calculated spin Nernst conductivity (SNC) of CoGe is also large, being -1.3 ($\hbar$/e)(A/m K) at room temperature.
Due to their semimetallic nature with low electrical conductivity, 
these topological semimetals may have large spin Hall and spin Nernst angles, being comparable to that of platinum metal.
The SHC and SNC of these compounds can also be increased by raising or lowering chemical potential
to, e.g., the topological nodes, via either chemical doping or electrical gating.
Our findings thus indicate that transition metal monosilicides of the CoSi family
not only would provide a material platform for exploring novel spin-transports and exotic phenomena 
in unconventional chiral fermion semimetals but also could be promising
materials for developing better spintronic and spin caloritronic devices.
	
\end{abstract}

\maketitle

\section{INTRODUCTION}

\begin{figure}[htbp] \centering 
\includegraphics[width=8.6cm]{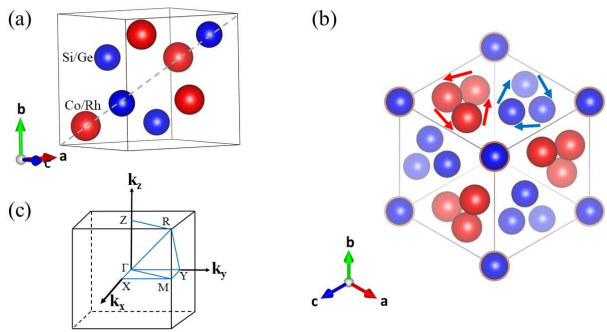}
\caption{Cubic B20 chiral crystal structure of the CoSi family. (a) Cubic primitive unit cell.
(b) Top view along the [111] direction [the dashed grey line in (a)] of the crystal structure 
(2$\times2\times$2 supercell).
Here the transparency of the atoms denotes the depth of the atomic positions from top to bottom. 
The red and blue arrows indicate the right-handed and left-handed helicity (chirality) 
of the Co/Rh and Si/Ge atoms, respectively. In this paper, the helicity of the Co/Rh atoms 
is used to define the chirality of the crystal because the energy bands near the Fermi level
are dominated by Co/Rh $d$ orbitals. (c) The corresponding cubic Brillouin zone.}
\label{fig:crystal}
\end{figure}

Spin current generation and manipulation are key issues in spintronics.
Spin Hall effect (SHE) \cite{Dyakonov1971,Hirsch1999,Murakami2003,Sinova2004,Kato2004,Guo2005,Valenzuela2006,Chang2007,Kimura2007,Guo2008,Tanaka2008,Liu2012,Hoffmann2013,Sinova2015}, 
i.e., generation of pure transverse spin current in a nonmagnetic material with relativistic electron interaction
(spin-orbit coupling) by an electric field, was first proposed by Dyakonov and Perel in 1971 \cite{Dyakonov1971}. 
It requires neither an applied magnetic field nor a magnetic material to produce a pure spin current
\cite{Hirsch1999,Murakami2003,Sinova2004,Kato2004,Guo2005,Valenzuela2006,Chang2007,Kimura2007,Guo2008,Tanaka2008},
thus offering an important advantage for the fabrication of low-power-consumption spintronic devices~\cite{Hoffmann2013,Sinova2015}
such as spin-orbit torque switching based nanodevices \cite{Liu2012}.       
Large intrinsic SHE has been predicted and also observed in several 5$d$ transition metals such as 
platinum because of their strong spin-orbit coupling (SOC) \cite{Kimura2007,Guo2008,Tanaka2008,Liu2012,Hoffmann2013,Sinova2015}. 
More recently, it was predicted that transverse spin current could also be generated in a nonmagnetic material 
by applying a temperature gradient ($\nabla T$) instead of an electric field $E$ \cite{Cheng2008}.
This thermoelectric analog of SHE is known as spin Nernst effect (SNE) and
would make spintronic devices powered by heat possible, leading to a new field called spin caloritronics \cite{Bauer2012}.
Large SNE was recently observed in platinum and tungsten metals.~\cite{Meyer2017,Sheng2017}

In the past few years, the study of SHE  
\cite{Okuma2016,Sun2016,Sun2017,MacNeill2017,Shi2019,Zhao2020,Yen2020,Prasad2020,Ng2021,Tang2021b} and 
SNE \cite{Yen2020,Prasad2020} in so-called topological semimetals has attracted considerable attention.
In high-energy physics, the standard model predicts three kinds of fermionic particles in the
Universe, namely, Dirac, Weyl and Majorana fermions, on the basis of the Poincare group.
However, only Dirac fermions have been captured so far.
Interestingly, in condensed matter physics, a variety of fermionic quasiparticles have been realized
in topological semimetals which are not constrained by the Poincare symmetry \cite{Bradlyn2016,Chang2017,Tang2017,Armitage2018}. 
In a Dirac semimetal, which is usually a nonmagnetic centrosymmetric crystal,
the bulk band structure hosts four-fold degenerate band crossing points (called Dirac points) with linearly dispersed
excitations described by the $4\times4$ Dirac Hamiltonian~\cite{Armitage2018}. 
When the spatial inversion symmetry is broken, as in a Weyl semimetal (WSM), a Dirac point
is split and produces a pair of two-fold stable band crossing points (called Weyl points)
with linearly dispersed excitations described by the $2\times2$ Weyl Hamiltonian~\cite{Armitage2018}. 
A pair of Weyl points behave as a pair of monopoles of Berry curvature in momentum space 
and carry opposite chiral charges (Chern numbers) of $\pm$1 ~\cite{Armitage2018}. 
Since the SHE could be considered as derived from the interplay of the spin-momentum locking and 
large Berry curvature of the electronic bands near the Weyl points~\cite{Sinova2015,Xiao2010}, one could expect
large intrinsic SHE in Weyl semimetals (WSMs)~\cite{Sun2016}. 
Indeed, the TaAs family of Weyl semimetals (WSMs), was predicted to exhibit large SHE \cite{Sun2016}. 
More recently, large SHE was observed in WSM WTe$_{2}$ \cite{MacNeill2017,Shi2019,Zhao2020}.

Transition metal monosilicides CoSi, CoGe, RhSi and RhGe (known as the CoSi family)
crystallize in a chiral cubic lattice \cite{Kavich1978,Demchenko2008,Takizawa1988,Engstrom1965,Larchev1982} (see Fig. 1).
Interestingly, new types of chiral fermions beyond spin-1/2 Weyl fermions, such as spin-3/2 and spin-1 
chiral fermions, have recently been discovered in structurally chiral crystals including the 
CoSi family considered here.~\cite{Bradlyn2016,Chang2017,Tang2017,Chang2018}
Unlike spin-1/2 Weyl fermions, spin-3/2 and spin-1 fermionic quasiparticles have no counterpart
in high-energy physics, and thus are called unconventional (or multifold) chiral fermions.
Unlike Weyl points, multifold chiral fermion nodes sit on high symmetry points and lines in the
Brillouin zone with their chiral charges being larger than $\pm$1. 
Furthermore, two partners of a pair of nodal points can be located at 
two different energy levels.~\cite{Bradlyn2016,Chang2017,Tang2017,Chang2018} 
As a result, unconventional chiral fermion semimetals were predicted to 
exhibit exotic physical phenomena 
such as long Fermi arc surface states \cite{Chang2017,Tang2017,Rao2019},
gyrotropic magnetic effect \cite{Zhong2016} and quantized circular photogalvanic effect \cite{Juan2017}. 

Therefore, we may expect that the CoSi family would exhibit novel spin transport phenomena and
thus become useful materials for spintronic and spin caloritronic devices.
In nonchiral cubic crystals such as platinum and tungsten metals, the spin Hall conductivity (SHC) tensor has      
only one independent nonzero element ($\sigma_{xy}^z$), and $\sigma_{yx}^z = -\sigma_{xy}^z$~\cite{Guo2008}.
In contrast, because of the absence of the chiral symmetry, the SHC tensor of the CoSi family 
has two independent nonzero elements and in general $\sigma_{yx}^z = \sigma_{xz}^y \ne -\sigma_{xy}^z$, 
as will be reported below in Sec. III(B).
Among the CoSi family, only the SHE in CoSi was recently studied by combining experiments and first-principles calculations,
and the SHC ($\sigma_{xy}^z$) was found to be quite large.~\cite{Tang2021b}
However, the other SHC element $\sigma_{xz}^y$ was not considered in Ref. \cite{Tang2021b}.
As will be reported below in Sec. III(B), the knowledge of both independent nonzero
SHC tensor elements would allow us to determine the helicity of the chiral fermions 
in the CoSi family and also to identify their structural chirality.
Note that detection of the relationship between chirality of chiral fermions and structural chiral crystals
has currently attracted considerable interest~\cite{Ma2017,Li2019,Rees2020,Sun2020}.

No study of the SNE in the CoSi family has been reported. 
We notice that members of the CoSi family are semimetals with large Seebeck coefficient and thermopower
and thus have been extensively studied as thermoelectric materials for many years~\cite{Asanabe1964,Skoug2009,Kanazawa2012,Sidorov2018}.
For example, CoSi and CoGe have large negative Seebeck coefficent of about -80 $\mu$V/K~\cite{Skoug2009,Kanazawa2012}.
Thus, one could expect the CoSi family to be exploited for thermal spin current generation via SNE. 
Furthermore, recent first-principles calculations~\cite{Tang2021b} showed that the energy derivative of the SHC $\sigma_{xy}^z(E_F)'$
at the Fermi level ($E_F)$ in CoSi is very large. This further suggests that large SNE could occur in the CoSi family
because the Mott relation [see Eq. (4) below] says that the spin Nernst conductivity (SNC) 
is proportional to $\sigma_{xy}^z(E_F)'$~\cite{Xiao2010,Yen2020}.

In this article, therefore, we present a systematic study of the SHE and SNE as well as topological aspects of
band structure of these unconventional chiral fermion compounds CoSi, CoGe, RhSi, and RhGe by performing \textit{ab initio} 
density functional theory (DFT) calculations.  The rest of this article is organized as follows.  
In Sec. II, we introduce the crystal structure of the CoSi family, followed by a brief description 
of the (spin) Berry phase formalism for calculating the intrinsic spin Hall conductivity 
and spin Nernst conductivity (SNC) along with the computational details.
The main results are presented in Sec. III which consists of four subsections.  In Sec. III A, we first analyze
the topological properties of the calculated relativistic band structures and density of states of the CoSi family.
Calculated SHC and SNC for the CoSi family are presented in Secs. III B and III C, respectively,
where we also compare our results with that in other known materials.
Finally, an analysis of the $k$-resolved spin Berry curvature is presented in Sec. III D in order to understand 
the origins of the calculated intrinsic SHC and SNC of the CoSi family. 
Finally, we summarize the conclusions drawn from this work in Sec. IV.

\section{THEORY AND COMPUTATIONAL DETAILS}

The CoSi family crystalizes in the simple cubic {\it B}$20$-type structure with space 
group {\it P}$2_{1}3$ (see Fig. 1).~\cite{Kavich1978,Demchenko2008} 
As mentioned before, this structure is structurally chiral, and when viewed along the [111] axis, 
it can be either a right-handed crystal (RHC) or a left-handed crystal (LHC) [see Fig. 1(b)]. 
Nevertheless, the structural chirality of a grown crystal could not be pre-specified and it would depend
perhaps on the specific growth process. Interestingly, based on our calculated SHCs to be presented in the next section,
we find that the crystal structures of CoSi reported, respectively, in Refs. ~\cite{Kavich1978} and ~\cite{Demchenko2008}
have the opposite chiralities. Nevertheless, both RHC and LHC structures have the same band dispersions.
Furthermore, as will be reported in Sec. III, their SHC and SNC tensors are related 
although they are different. Therefore, in this study, we focus on the right-handed crystals~\cite{Kavich1978}
unless stated otherwise.
In the present {\it ab initio} calculations, we use the experimental lattice constants
as well as measured atomic positions for all the four considered compounds.\cite{Kavich1978,Takizawa1988,Engstrom1965,Larchev1982}

Our self-consistent electronic structure calculations are based on the density functional theory (DFT) 
with the generalized gradient approximation (GGA) \cite{Perdew1996}. 
The accurate projector augmented-wave method \cite{Blochl1994}, 
as implemented in the Vienna Ab Initio Simulation Package (VASP) \cite{Kresse1993,Kresse1996}, is used.
The valence electronic configurations of Co, Rh, Si, and Ge taken into account in the present 
{\it ab initio} study are $3d^84s^1$,  $4d^85s^1$, $3s^23p^2$, and $3d^{10}4s^24p^2$, respectively.
A large plane wave energy cutoff of 400 eV is used throughout.
In the self-consistent electronic structure calculations, a $\Gamma$-centered $k$-point 
mesh of 16 $\times$ 16 $\times$ 16 is used in the Brillouin zone (BZ) integration 
by the tetrahedron method \cite{Jepson1971,Temmerman1989}.
However, for the density of states (DOS) calculations, a denser $k$-point grid 
of 24 $\times$ 24 $\times$ 24 is adopted.

The intrinsic SHC is calculated via the Kubo formula in the clean limit ($\omega = 0$) 
within the elegant Berry-phase formalism \cite{Guo2008,Xiao2010,Guo2017}.
Within this formalism, the SHC ($\sigma_{ij}^s=J_i^s/E_j$) 
is given by the BZ integration of the spin Berry curvature for all 
the occupied bands below the Fermi level $E_{F}$~\cite{Guo2008},
\begin{equation}
\label{eq:1}
\begin{aligned}
\sigma_{ij}^{s}= e\sum_{n}\int_{BZ}\frac{d\textbf{k}}{(2\pi)^3}f_{\textbf{k}n}\Omega_{ij}^{n,s}(\textbf{k}),
\end{aligned}
\end{equation}

\begin{equation}
\label{eq:2} 
\begin{aligned}
\Omega_{ij}^{n,s}{\textbf{(k)}}=\sum_{n'\neq n}\frac{2Im\left[\langle\textbf{k}n|\{\tau_{s},v_{i}\}/4|\textbf{k}n'\rangle\langle\textbf{k}n'|v_{j}|\textbf{k}n\rangle \right] }{(\epsilon_{\textbf{k}n}-\epsilon_{\textbf{k}n'})^2+(\eta)^2},
\end{aligned}
\end{equation}  
where $f_{\textbf{k}n}$ is the Fermi distribution function, and $\Omega_{ij}^{n,s}{\textbf{(k)}}$ 
is the spin Berry curvature for the $n^{th}$ band at $\textbf{k}$ with $i,j\in (x,y,z)$ and $i\neq j$. 
$\tau_s$, $v_i$ ($v_j$), and $\eta$ denote the Pauli matrix, velocity operator, 
and fixed smearing parameter, respectively. $J_i^s$ is the $i$th component of the spin current density $J^s$, $E_j$ 
is the $j$th component of the electric field $E$, and $s$ is the spin direction, respectively. 
Once the SHC is calculated, the SNC ($\alpha_{ij}^s=-{J_{i}^{s}}/{\nabla_{j}T}$)
is obtained by an energy integration of the calculated SHC \cite{Xiao2010},
\begin{equation}
\label{eq:3}
\begin{aligned} 
\alpha_{i j}^{s}=\frac{1}{e} \int_{-\infty}^{\infty} d \varepsilon \frac{\partial f}{\partial \varepsilon} \sigma_{i j}^{s}(\varepsilon) \frac{\varepsilon-\mu}{T}
\end{aligned}
\end{equation}
where $\mu$ is the chemical potential.

Since a large number of $k$ points are required to get accurate SHC and SNC, 
we use the efficient Wannier function interpolation scheme \cite{Qiao2018,Ryoo2019} based on 
the maximally localized Wannier functions (MLWFs) \cite{Marzari2012} as implemented 
in the WANNIER90 package \cite{Pizzi2020}.
Since the energy bands around the Fermi level are dominated by mainly transition metal $d_{xy}$ orbitals, 
8 $d_{xy}$ orbital MLWFs per unit cell of the CoSi family are constructed by fitting to 
the \textit{ab initio} relativistic band structure 
in the energy window from -0.5 eV to 0.5 eV around $E_{F}$.
The band structure obtained by the Wannier interpolation for the CoSi family agrees well 
with that from the \textit{ab initio} calculation as can be seen in Fig. S1 in the Supplemental Material (SM) \cite{Prasad-SM}. 
The SHC for all the four compounds of the CoSi family is then evaluated by taking 
a very dense $k$ mesh of 200 $\times$ 200 $\times$ 200, 
with a 5 $\times$ 5 $\times$ 5 adaptive refinement scheme.
Test calculations using several different sets of $k$ meshes show that the calculated SHC and SNC 
for all four considered compounds of the CoSi family converge within a few percent.

\section{RESULTS AND DISCUSSION}

\subsection{Electronic structure}

\begin{figure}[htbp] \centering
\includegraphics[width=8.2cm]{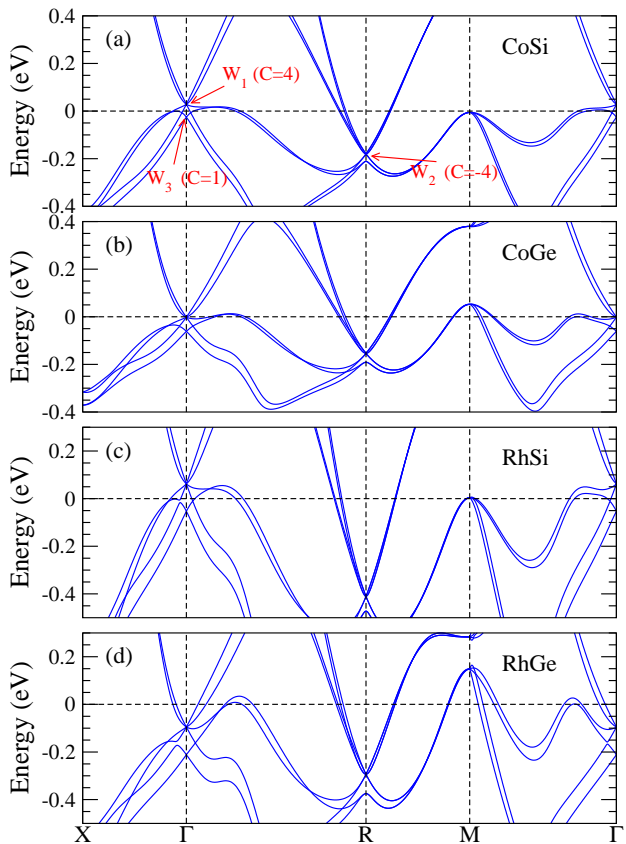}
\caption{Relativistic band structures of (a) CoSi, (b) CoGe, (c) RhSi, and (d) RhGe. 
Here the Fermi level is at the zero energy. In (a), the topological nodes
at the $\Gamma$ and $R$ points are labeled in red together with their Chern numbers
(chiral charges).}
\label{fig:band}
\end{figure}

\begin{figure}[htbp] \centering
\includegraphics[width=8.2cm]{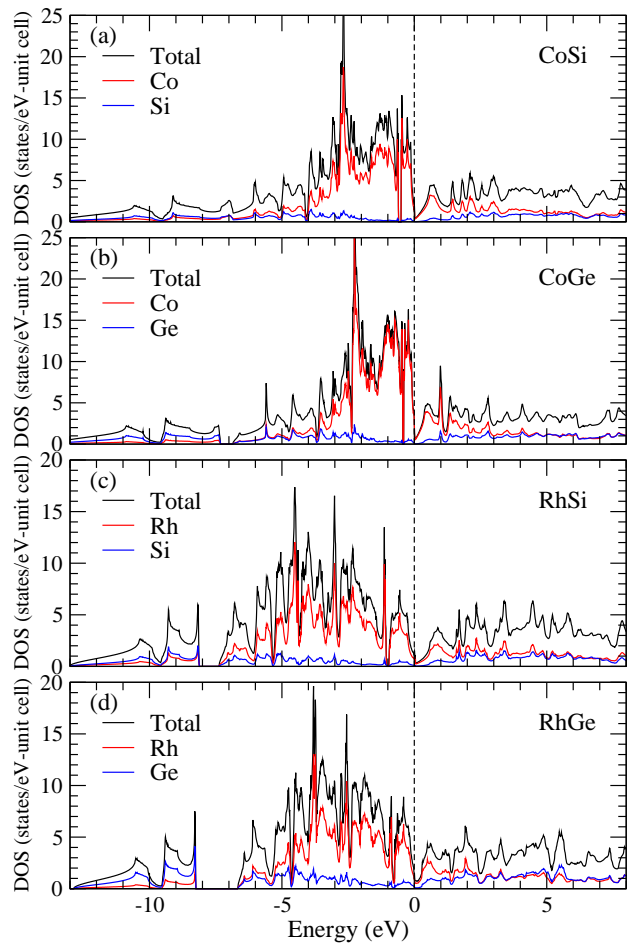}
\caption{Total and atom-decomposed density of states (DOS) of (a) CoSi, (b) CoGe, (c) RhSi, and (d) RhGe. 
Here the Fermi level is at the zero energy.}
\label{fig:dos}
\end{figure}

We find that all four considered compounds of the CoSi family possess a nonmagnetic ground state. 
The calculated relativistic band structures and density of states (DOS) of the CoSi family are shown 
in Figs. \ref{fig:band} and \ref{fig:dos}, respectively.
We also calculate the scalar-relativistic band structures of the CoSi family [i.e., without including  
the SOC], as shown in Fig. S2 in the SM \cite{Prasad-SM}.
Compared with scalar-relativistic bands (Fig. S2), relativistic bands (Fig. \ref{fig:band}) 
are splitted along the high symmetry points and lines in the BZ. 
In particular, at the $\Gamma$ point, the six-fold band crossing near $E_F$ in the scalar-relativistic 
band structures (Fig. S2) is split and becomes 
two band crossing points at two different energy levels with four-fold and two-fold degeneracy in the vicinity
of the Fermi level when the SOC is included (see Fig. 2 and also Figs. S3-S6 in the SM \cite{Prasad-SM}). 
The calculated Chern numbers of the four-fold (W$_1$) and two-fold (W$_3$) degenerate band crossing points
are $+4$ and $+1$ [see Fig. 2(a)], respectively. Therefore, nodal points W$_1$ and W$_3$
are right-handed chiral fermion nodes with chiral charges of $+4$ and $+1$, respectively. 
At the $R$ point, the eight-fold band crossing below $E_F$ in the scalar-relativistic band structures
(Fig. S2 in the SM \cite{Prasad-SM}) also split. This results in two band crossing points, i.e.,
one band crossing (W$_2$) with six-fold degeneracy having a Chern number of $-4$ 
(i.e., left-handed chiral fermion node) and the other band crossing below W$_2$ in the relativistic case (see Fig. 2).  
While the two-fold band crossing W$_3$ with Chern number of +1 is the conventional Weyl fermion node, 
the four-fold band crossing W$_1$ at the $\Gamma$ and six-fold band crossing W$_2$ at the $R$ point 
are known as spin-$3/2$ Rarita-Schwinger-Weyl (RSW) fermion and spin-$1$ double Weyl fermion nodes, 
respectively \cite{Bradlyn2016,Chang2017,Tang2017}.
Unlike Weyl fermions, spin-$3/2$ RSW fermion and spin-$1$ double Weyl fermions have no counterpart in high-energy physics. 
Therefore, they are often called unconventional (i.e., beyond the standard model) chiral fermions \cite{Bradlyn2016,Chang2017,Tang2017,Rao2019}. 
Interestingly, the nodes at the $\Gamma$ point are at close vicinity to $E_F$ (see Table S1 in the SM~\cite{Prasad-SM} for detailed values), and thus may significantly affect the spin transport properties such as SHC and SNC. 
The nodes at the $R$ point, however, lie further below $E_F$ for the investigated compounds, and thus their impact can be expected to be much smaller.
The origin of unconventional chiral fermions in the studied compounds can be understood 
by symmetry analysis as described in Refs. \cite{Chang2017} and \cite{Tang2017}.

Figure \ref{fig:dos} shows the total and atom-decomposed density of states (DOSs) 
as a function of energy ($E$) of the CoSi family. 
Total and atomic orbital-decomposed DOS spectra are also shown in Fig. S7 in the SM \cite{Prasad-SM}. 
All the four investigated compounds have a pseudo-gap at the Fermi level 
and thus have a low DOS value at $E_F$ (Fig. \ref{fig:dos} and Fig. S7), as can be expected from a semimetal.
Interestingly, in CoSi and CoGe, the DOS increases sharply as $E$ is lowered below $E_F$ 
but increases much less dramatically if $E$ is raised above $E_F$. This strong electron-hole
asymmetry in the DOS in the vicinity of $E_F$ was attributed to cause the large negative Seebeck
coefficient of about -80 $\mu$V/K in CoSi and CoGe.~\cite{Kanazawa2012}
From Fig. \ref{fig:dos} (as well as Fig. S7 in the SM \cite{Prasad-SM}), it is clear that Co (Rh) $d$ orbitals 
give the main contribution to the DOS across a wide energy range around $E_F$ with a minor contribution 
from Co (Rh) $p$ orbitals, and Si (Ge) $p$ orbitals. 
However, Si (Ge) $s$ orbitals are dominant for the lower energy bands below -7.4 eV (see Fig. \ref{fig:dos} and Fig. S7).   

\begin{table*}
\caption{\textit{P}2$_1$3 symmetry-imposed shape of the SHC tensor for the CoSi family in the right-handed (RHC) 
(left column) and left-handed (LHC-$M$) (right column) crystal.  
Here, the LHC-$M$ is obtained from the RHC via a nonsymmorphic operation involving mirror reflection $M$, e.g.,
($M_{1\bar{1}0}|\frac{3}{4}\frac{3}{4}\frac{3}{4}$) (see Table S2 in the SM ~\cite{Prasad-SM}).  
Clearly, $\sigma_{xy}^{z}$(LHC-$M$) $= -\sigma_{xz}^{y}$(RHC) and $\sigma_{xz}^{y}$(LHC-$M$) $= -\sigma_{xy}^{z}$(RHC). 
In contrast, for the LHC obtained from the RHC through a parity operation $\mathcal{P}$ (i.e., LHC-$\mathcal{P}$),
the nonzero elements of the SHC tensor are the same as that of the RHC (see the maintext).
Note that there are only two inequivalent nonzero elements (i.e., $\sigma_{xy}^{z}$ and $\sigma_{xz}^{y}$)
and that the shape of the SNC tensor is the same as that of the SHC.}
\begin{tabular}{c c c c c c}
\hline\hline
& SHC (RHC) & & & SHC (LHC-$M$) & \\
\underline{$\sigma$}$^x$ & \underline{$\sigma$}$^y$ & \underline{$\sigma$}$^z$ & \underline{$\sigma$}$^x$ & \underline{$\sigma$}$^y$ & \underline{$\sigma$}$^z$ \\
\hline
\\
$\left(\begin{array}{ccc} 0 & 0 & 0\\0 & 0 & \sigma_{xy}^{z}\\0 & \sigma_{xz}^{y}&0\end{array}\right)$ &
$\left(\begin{array}{ccc} 0 & 0 & \sigma_{xz}^{y}\\0 & 0 & 0\\\sigma_{xy}^{z} & 0&0\end{array}\right)$ &
$\left(\begin{array}{ccc} 0 & \sigma_{xy}^{z} & 0\\\sigma_{xz}^{y} & 0 & 0\\0 & 0&0\end{array}\right)$ &
$\left(\begin{array}{ccc} 0 & 0 & 0\\0 & 0 & -\sigma_{xz}^{y}\\0 & -\sigma_{xy}^{z}&0\end{array}\right)$ &
$\left(\begin{array}{ccc} 0 & 0 & -\sigma_{xy}^{z}\\0 & 0 & 0\\-\sigma_{xz}^{y} & 0&0\end{array}\right)$ &
$\left(\begin{array}{ccc} 0 & -\sigma_{xz}^{y} & 0\\-\sigma_{xy}^{z} & 0 & 0\\0 & 0&0\end{array}\right)$ \\ \\
\hline\hline
\end{tabular}
\label{table:1}
\end{table*}

\begin{table*}
\caption{
Calculated spin Hall conductivity ($\sigma_{xy}^{z}$ and $\sigma_{xz}^{y}$),
and spin Nernst conductivity ($\alpha_{xy}^{z}$ and $\alpha_{xz}^{y}$) at temperature $T=300$ K of the CoSi family.
Previous results for Weyl semimetal TaAs, Dirac line-node semimetal ZrSiS, and heavy 
Pt metal are also listed for comparison. To estimate spin Hall (Nernst) angle $\Theta_{sH} = 2\sigma^s/\sigma_{xx}^{c}$  
[$\Theta_{sN}= 2\alpha^s/\alpha_{xx} = 2\alpha^s/(S_{xx}\sigma_{xx}^{c})$], we also list experimental
electrical conductivity $\sigma_{xx}^{c}$ and Seebeck coefficient $S_{xx}$ values as well as
the estimated $\Theta_{sH}$ and $\Theta_{sN}$ here.
}
\begin{ruledtabular}
\begin{tabular}{c c c c c c c c c c c}
System & $\sigma_{xx}^{c}$ & $S_{xx}$ & $\sigma_{xy}^{z}$ & $\Theta_{sH}^{z}$ & $\sigma_{xz}^{y}$ & $\Theta_{sH}^{y}$ & $\alpha_{xy}^{z}$ & $\Theta_{sN}^{z}$ & $\alpha_{xz}^{y}$ & $\Theta_{sN}^{y}$ \\
& (S/cm) & ($\mu$V/K) & ($\hbar$/e)(S/cm) & (\%) & ($\hbar$/e)(S/cm) & (\%) & ($\hbar$/e)(A/m-K) & (\%) & ($\hbar$/e)(A/m-K) & (\%) \\ \hline
CoSi & 5200$^{c}$&-81$^{c}$ & -63, 52$^{j}$& -2.4& -66& -2.5 & 0.42 & -2.0 & -1.00 & 4.7\\
CoGe & 4589$^{d}$&-82$^{d}$ & -131         & -5.7& -21& -0.9 & 0.06 & -0.3 & -1.25 &  6.6 \\
RhSi & 3571$^{e}$&-25$^{h}$ & -122         & -6.8&  11&  0.6 & 0.14 & -3.1 & -0.65 &  14.6\\
RhGe & 4130$^{f}$&-25$^{f}$ & -139         & -6.7& 103&  5.0 &-0.64 &  12.4 &-0.19 &  3.7\\
TaAs$^{a}$ &  -- &-- &-781 & --& 357 & -- & -- & -- & -- & -- \\
ZrSiS$^{b}$& -- & -- &  79 & --& -280 & -- & 0.60 & -- &  0.52 & -- \\
Pt        & 20833$^{g}$ & -3.7$^{i}$ & 2139$^{k}$ & 10$^{g}$ & -- & -- & -1.09$^{m}$,-1.57$^{i}$ & -20$^{i}$ & -- & -- \\
\end{tabular}
\end{ruledtabular}
{$^{a}$\textit{Ab initio} calculation \cite{Sun2016};
 $^{b}$\textit{Ab initio} calculation \cite{Yen2020};
 $^{c}$Transport experiment \cite{Skoug2009};
 $^{d}$Transport experiment \cite{Kanazawa2012};
 $^{e}$Transport experiment \cite{Maulana2020};
 $^{f}$Transport experiment \cite{Sidorov2018};
 $^{g}$Transport experiment \cite{Wang2014};
 $^{h}$Assumed the same value as RhGe from \cite{Sidorov2018};
 $^{i}$ Experiment at 255 K \cite{Meyer2017};
 $^{j}$\textit{Ab initio} calculation \cite{Tang2021b};
 $^{k}$\textit{Ab initio} calculation \cite{Guo2008};
 $^{m}$\textit{Ab initio} calculation \cite{Guo2017}.}
\label{table:2}
\end{table*}

\subsection{Spin Hall effect}

\begin{figure}[htbp] \centering
\includegraphics[width=8.0cm]{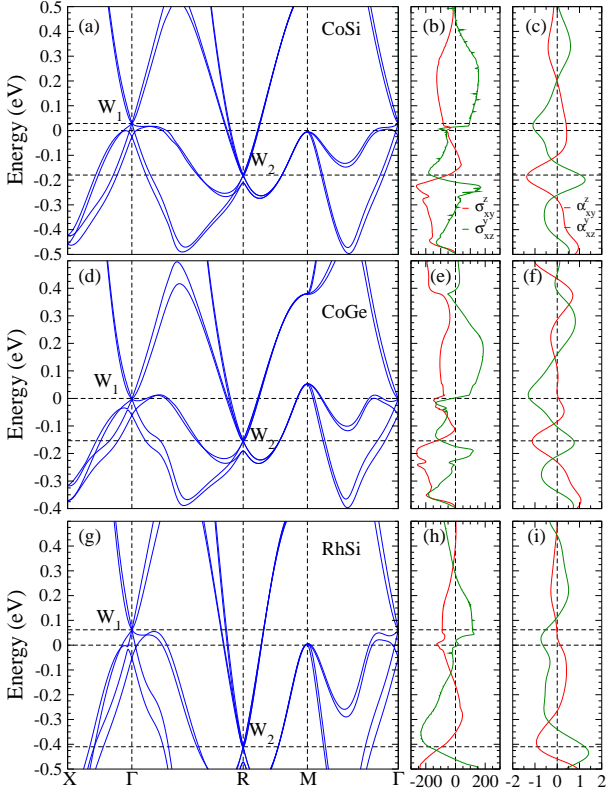}
\caption{(a, d, g) Relativistic band structure,
(b, e, h) spin Hall conductivity ($\sigma_{xy}^{z}$ and $\sigma_{xz}^{y}$) as a function of chemical potential ($\mu$),
and (c, f, i) spin Nernst conductivity ($\alpha_{xy}^{z}$ and $\alpha_{xz}^{y}$) at $T=300$ K 
as a function of $\mu$ for CoSi (a, b, c), CoGe (d, e, f) and RhSi (g, h, i).
Here W$_{1}$ and W$_{2}$ denote spin-3/2 chiral fermion node at $\Gamma$ and spin-1 double Weyl node at $R$, respectively. 
The Fermi level is at the zero energy, and the unit of the SHC (SNC) is ($\hbar$/e)(S/cm) [($\hbar$/e)(A/m K)].}
\label{fig:SHC-XY}
\end{figure}

\begin{figure}[htbp] \centering
\includegraphics[width=8.0cm]{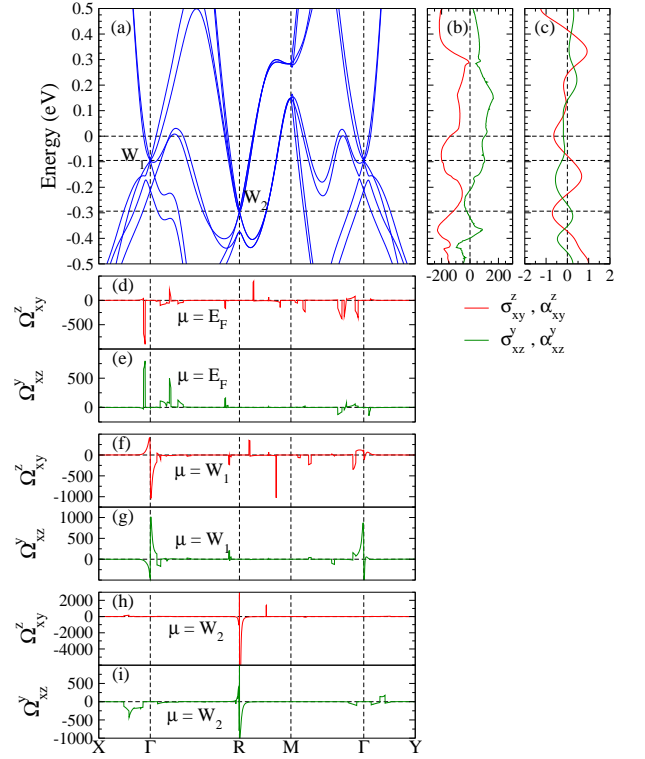}
\caption{RhGe. (a) Relativistic band structure;
(b) spin Hall conductivity ($\sigma_{xy}^{z}$ and $\sigma_{xz}^{y}$) as a function of chemical potential ($\mu$);
(c) spin Nernst conductivity ($\alpha_{xy}^{z}$ and $\alpha_{xz}^{y}$) at $T=300$ K as a function of $\mu$;
(d), (f), and (h) spin Berry curvature (SBC) $\Omega_{xy}^{z}$;
as well as (e), (g), and (i) SBC $\Omega_{xz}^{y}$ 
for $\mu$ = $E_{F}$, $\mu$ = W$_{1}$, and $\mu$ = W$_{2}$, respectively, along the 
high symmetry lines in the Brillouin zone. 
Here W$_{1}$ and W$_{2}$ denote spin-3/2 chiral fermion node 
at $\Gamma$ and spin-1 double Weyl node at $R$, respectively.
In (a), (b), and (c), the Fermi level is at the zero energy,
and the unit of the SHC (SNC) is ($\hbar$/e)(S/cm) [($\hbar$/e)(A/m K)].
In (d) - (i), the unit of SBC is \AA$^{2}$.}
\label{fig:RhGe}
\end{figure}

From Eq. (\ref{eq:1}), it is clear that the SHC of a material is a third-rank tensor ($\sigma_{ij}^{s}$; $s,i,j=x,y,z$),
and thus has 27 tensor elements in total. Nevertheless, due to the constraint of the crystalline symmetry of the material,
many of these tensor elements become zero. \cite{Gallego2019}
Indeed, due to the high symmetry of the cubic {\it B}$20$-type structure, the CoSi family have only 
two independent nonzero tensor elements, namely, $\sigma_{xy}^{z}$ and $\sigma_{yx}^{z}$. \cite{Gallego2019} 
Other nonzero elements are related to these two elements by $ \sigma_{yz}^{x} = \sigma_{zx}^{y} = \sigma_{xy}^{z}$ 
and $\sigma_{xz}^{y} = \sigma_{zy}^{x} = \sigma_{yx}^{z} $ (see Table I).
Note that for nonchiral cubic materials such as Pt metal~\cite{Guo2008}, 
there is only one independent nonzero element, i.e., $\sigma_{xy}^{z}$, and $\sigma_{yx}^{z}=-\sigma_{xy}^{z}$. 
The CoSi family have two independent nonzero tensor elements
because their structural chiral symmetry is broken, as mentioned above in Sec. II.
Therefore, the magnitude of ($\sigma_{xy}^{z}$ + $\sigma_{yx}^{z}$) is a measure of the structural chirality of the CoSi family.
Following the convention \cite{Gallego2019}, we will focus on $\sigma_{xy}^{z}$ and $\sigma_{xz}^{y}$
instead of $\sigma_{xy}^{z}$ and $\sigma_{yx}^{z}$ below.
In Table II, we list the calculated independent nonzero elements of the SHC tensor for the CoSi family.
The $\sigma_{xy}^{z}$ and $\sigma_{xz}^{y}$ of recently studied Weyl semimetal TaAs \cite{Sun2016} and Dirac line-node semimetal ZrSiS \cite{Yen2020} 
as well as the $\sigma_{xy}^{z}$ of platinum metal \cite{Guo2008,Meyer2017}, are listed also in Table II for comparison.

It is well known that the handedness of a chiral crystal can be transformed simply by a spatial inversion (parity) operation ($\mathcal{P}$),
e.g., from the RHC to LHC. Interestingly, here we discover that there are many other operations which can also change
the handedness of the chiral crystal from the RHC to LHC or vice-versa. For example, for the CoSi family, we find that there are
nine nonsymmorphic glide operations, each consisting of a mirror reflection and a fractional lattice translation 
[e.g., ($M_{1\bar{1}0}|\frac{3}{4}\frac{3}{4}\frac{3}{4}$)], which would transform the handedness from the RHC to LHC,
as listed in Table S2 in the SM \cite{Prasad-SM}. For clarity, let us label the LHC obtained from the RHC through a parity operation
LHC-$\mathcal{P}$, and the LHC resulted from the RHC via a nonsymmorphic operation involving a mirror reflection LHC-$M$.
Furthermore, we find that the SHC tensor elements for the RHC and LHC-$\mathcal{P}$ crystals are the same. 
This can be seen as follows. Equation (1) shows that the SHC is given by the sum of spin Berry curvature 
over the BZ. For a system where the spin is a good quantum number, 
$\Omega_{ij}^{n,k} = s_k \Omega_{ij}$.~\cite{Sun2016} Thus, the spin Berry curvatures and hence the SHC 
would remain unchanged under parity because both the Berry curvatures ($\Omega_{xy}$ and $\Omega_{xz}$) 
and spin operators ($S^{z}$ and $S^{y}$) are pseudovectors and thus even under parity.
To verify this conclusion, we perform explicit calculations of the SHC for both RHC and LHC-$\mathcal{P}$ of CoSi,
and indeed find that the calculated $\sigma_{xy}^{z}$ and $\sigma_{xz}^{y}$ values 
for these two structures are identical (within the numerical uncertainties).

Surprisingly, we find that the SHC tensor elements for the RHC and LHC-$M$ structures are different.  
Nevertheless, they are connected via simple relations $\sigma_{xz}^{y}$(LHC-$M$) $= -\sigma_{xy}^{z}$(RHC) 
and $\sigma_{xy}^{z}$(LHC-$M$) $= -\sigma_{xz}^{y}$(RHC) (see Table I). 
To understand the origin of these SHC relations, let us take the nonsymmorphic 
operation ($M_{1\bar{1}0}|\frac{3}{4}\frac{3}{4}\frac{3}{4}$) as an example.
As the mirror reflection ($M_{1\bar{1}0}$) changes the Berry curvature $\Omega_{xz}$ to $\Omega_{yz}$ 
and spin operator $S^{y}$ to -$S^{x}$, we get relation $\Omega_{xz}^{y}$(RHC) $= -\Omega_{yz}^{x}$(LHC-$M$) 
for spin Berry curvature. This results in $\sigma_{xz}^{y}$(RHC) $= -\sigma_{yz}^{x}$(LHC-$M$) $= -\sigma_{xy}^{z}$(LHC-$M$),
i.e., $\sigma_{xy}^{z}$(LHC-$M$)$= -\sigma_{xz}^{y}$(RHC), as mentioned earlier. 
We notice that the crystal structures of CoSi reported in Refs. ~\cite{Kavich1978} and ~\cite{Demchenko2008}
have the opposite chiralities. Furthermore, the structure reported by Demchenko {\it et al.} \cite{Demchenko2008}  
is nearly the same as the LHC-$M$ of the crystal structure reported by Kavich {\it et al.} \cite{Kavich1978} 
(with a measure of similarity $\Delta = 0.005$ \cite{Flor2016}) (see also Table S3 in the SM~\cite{Prasad-SM}).
Therefore, to verify these relations, we again perform explicit calculations of the SHC
for these structures reported by ~\cite{Kavich1978} (RHC) and ~\cite{Demchenko2008} (LHC-$M$), and the results are shown in Fig. S10.
Figure S10 shows that indeed $\sigma_{xy}^{z}$(LHC-$M$) $\approx$ -$\sigma_{xz}^{y}$(RHC)
and $\sigma_{xz}^{y}$(LHC-$M$) $\approx$ -$\sigma_{xy}^{z}$(RHC).
The small discrepancies are due to the slight differences in the atomic positions in these crystalline 
structures~\cite{Kavich1978,Demchenko2008} (see also Table S3). This shows that the SHC can be used to identify 
the handedness (chirality) for the crystal structure of the CoSi family as well as other chiral materials.
Here we mention two cases to demonstrate how measuring the SHC tensor elements can help determine the handedness 
(chirality) of a structurally chiral crystal. First, let us assume that both the SHC tensor elements ($\sigma_{xy}^{z}$ and $\sigma_{xz}^{y}$) 
are of the same sign, e.g., negative as in the case of RHC CoSi and CoGe at the Fermi level (see Table II). 
Now if one observes a positive sign for these two elements for the same crystal with unknown handedness, 
using our relations as shown in Table I for the RHC and LHC-$M$, one can conclude that the unknown handedness crystal is LHC-$M$. 
Second, when both the SHC tensor elements ($\sigma_{xy}^{z}$ and $\sigma_{xz}^{y}$) are of the opposite sign 
[as in the case of RhSi and RhGe at the Fermi level (see Table II); RHC], one needs to consider the magnitude 
of both SHC tensor elements as well to determine the handedness of the same crystal. Note that the two SHC tensor elements 
of RhSi and RhGe differ quite a lot in terms of magnitude and can be used to determine the handedness.
Furthermore, since the chirality [or the sign of the Chern number (chiral charge)] of a chiral fermion node
is determined by the structural chirality (see Table S1 in the SM), the calculated SHC can also be used to 
identify the helicity (chirality) of the chiral fermions in the CoSi family and also in other chiral lattices.
Note that detection of the relationship between chirality of chiral fermions and structurally chiral crystals 
is a topic of considerable current interest~\cite{Ma2017,Li2019,Rees2020,Sun2020}. 
Nevertheless, as mentioned above, we will focus on the CoSi family in the RHC structure in the following.

The calculated SHC values for the four investigated compounds are listed 
in the increasing order of their SOC strength in Table II.
The calculated $\sigma_{xy}^{z}$ for CoSi is the smallest and that of RhGe is the largest, 
a trend which apparently follows the increasing order of the SOC in the CoSi family.
Nevertheless, an even more pronounced trend is present in the calculated $\sigma_{xz}^{y}$ values. 
Furthermore, the sign of $\sigma_{xz}^{y}$ for RhSi and RhGe is opposite to that for CoSi and CoGe.
RhGe has the largest $\sigma_{xy}^{z}$ of -139 ($\hbar$/e)(S/cm), being larger than
that [79 ($\hbar$/e)(S/cm)] of Dirac line-node semimetal ZrSiS~\cite{Yen2020} but significantly
smaller than that [-781 ($\hbar$/e)(S/cm)] of archetypal Weyl semimetal TaAs~\cite{Sun2016}.
Nevertheless, the $\sigma_{xy}^{z}$ of RhGe is about $20$ times larger than the $\sigma_{xy}^{z}$ [7 ($\hbar$/e)(S/cm)] 
of Weyl semimetal NbP \cite{Sun2016}. 
Finally, the $\sigma_{xy}^{z}$ value of RhGe is about 15 times smaller than that
of platinum metal which possess the largest SHC among transition metals. 

Table II also shows a strong anisotropy of the SHC in the CoSi family.
By interchanging the applied electric field and spin polarization directions 
simultaneously from ($z$, $y$) to ($y$, $z$),
one can obtain the multifold enhancement of the SHC. 
Here, the first and second index in the parentheses correspond to 
the applied electric field and spin polarization directions, respectively.
For example, 
the $\sigma_{xy}^{z}$ of CoGe and RhSi 
is about $6$ times and $11$ times larger than 
the corresponding $\sigma_{xz}^{y}$, respectively.
Furthermore, the signs of the SHC $\sigma_{xy}^{z}$ and $\sigma_{xz}^{y}$ for RhSi and RhGe 
also differ.

Since chiral fermion nodes W$_1$ at the $\Gamma$ point and W$_2$ at the $R$ point do not lie exactly at $E_F$ 
(see Fig. \ref{fig:band} and Table S1 in the SM \cite{Prasad-SM}), we also calculate the SHC as 
a function of chemical potential ($\mu$) within the rigid-band approximation.
In the rigid-band approximation, we vary only the $\mu$ while keeping the band structure fixed.
The calculated SHC spectra are displayed in Figs. \ref{fig:SHC-XY}(b), \ref{fig:SHC-XY}(e), \ref{fig:SHC-XY}(h), 
and \ref{fig:RhGe}(b) for CoSi, CoGe, RhSi and RhGe, respectively.
In the supplementary Table S1, we also tabulate the SHC values 
when the chemical potential is shifted to either node W$_1$ at $\Gamma$ or node W$_2$ at $R$. 
It is clear that the SHC spectra for the CoSi family show a strong dependence on $\mu$. 
In particular, Fig. \ref{fig:SHC-XY}(b) shows that the magnitude of $\sigma_{xz}^{y}$ of CoSi  
decreases steeply and then changes sign as $\mu$ is raised from $E_F$ to 0.015 eV.
The positive $\sigma_{xz}^{y}$ then increases steeply as $\mu$ further increases until
$\mu$ is above node W$_1$. 
As suggested earlier in Ref. \cite{Tang2021b}, this rapid change of the magnitude and also the sign of $\sigma_{xz}^{y}$ 
could be a characteristic behavior of the SHC in the vicinity of a spin-3/2 chiral fermion node. 
Indeed, this pronounced behavior also shows up in the $\sigma_{xz}^{y}$ spectrum of CoGe around 
the spin-3/2 chiral fermion node W$_1$ at $\Gamma$ [see Fig. \ref{fig:SHC-XY}(e)]. 
This rapid change of $\sigma_{xz}^{y}$ in CoSi and CoGe can also be seen near spin-1 double Weyl 
fermion node W$_2$ at $R$ except the sign change occur well below W$_2$ (about 0.25 eV lower). 
Nevertheless, the $\sigma_{xz}^{y}$ spectrum of RhGe is rather flat in the vicinity of W$_1$ at $\Gamma$ [see Fig. \ref{fig:RhGe}(b)].
Furthermore, this steep slope and sign change of $\sigma_{xz}^{y}$ around W$_1$ is absent in the $\sigma_{xy}^{z}$ spectrum of the CoSi family.
Consequently, this indicates that the pronounced feature of $\sigma_{xz}^{y}$ found near node W$_1$ 
in CoSi and CoGe may not be universal for all spin-3/2 chiral fermions.

Other prominent changes due to the variation of the $\mu$ are as follows.
The $\sigma_{xz}^{y}$ of RhSi can be increased by a factor of about $10$ 
by slightly increasing the $\mu$ from $E_F$ to node W$_1$ at 0.06 eV 
[see Fig. \ref{fig:SHC-XY}(h) and Table S1 in the SM \cite{Prasad-SM}]. 
This can be easily realised via electron doping of merely 0.02 e/f.u. 
Also, $\sigma_{xz}^{y}$ of CoGe could be enhanced 
from -21 ($\hbar$/e)(S/cm) to -98 ($\hbar$/e)(S/cm) 
by lowering the $\mu$ to nodal point W$_2$ at -0.15 eV [see Fig. \ref{fig:SHC-XY}(e) and 
also Table S1 in the SM \cite{Prasad-SM}]. 
This can be achieved via hole doping of 0.27 e/f.u. 
Furthermore, $\sigma_{xy}^{z}$ of CoGe and RhGe could reach a very large value of -260 ($\hbar$/e)(S/cm) 
and -202 ($\hbar$/e)(S/cm) by lowering the chemical potential to -0.19 eV and -0.07 eV via hole doping 
of 0.38 e/f.u. and 0.04 e/f.u., respectively [see Figs. \ref{fig:SHC-XY}(e) and \ref{fig:RhGe}(b)]. 
The above discussion clearly suggests that the SHC of the CoSi family could be tuned in both magnitude
and sign by shifting the chemical potential to either the topological nodal points
or other energy levels, and this could be accomplished via either chemical doping or electrical gating.   

Among the members of the CoSi family, only the SHE in CoSi was recently studied experimentally 
using CoSi/CoFeB/MgO heterostructures by Tang {\it et al.} \cite{Tang2021b} The measured
values of the so-called dampinglike ($\sigma_{DL}$) and fieldlike ($\sigma_{FL}$) 
SHC for film thickness $t_{CoSi} = 7.2$ nm are
$\sigma_{DL} = 45$ ($\hbar$/e)(S/cm) and $\sigma_{FL} = 95$ ($\hbar$/e)(S/cm), being in quite 
good agreement with our theoretical $\sigma_{xy}^{z}$ and $\sigma_{xz}^{y}$ values listed in Table II. 
Tang {\it et al.} also performed DFT calculations for SHC $\sigma_{xy}^{z}$ of CoSi
and their calculated $\sigma_{xy}^{z}$ at $E_F$ is 52 ($\hbar$/e)(S/cm), agree very well
with our $\sigma_{xy}^{z}$ and $\sigma_{xz}^{y}$ values (Table II).
Interestingly, their calculated $\sigma_{xy}^{z}$ spectrum in the vicinity of $E_F$ 
(see Fig. 5(c) in \cite{Tang2021b}) looks rather similar to our $\sigma_{xz}^{y}$ spectrum 
[see Fig. \ref{fig:SHC-XY}(b)], except that the two have opposite signs.
In other words,  $\sigma_{xy}^{z}$ \cite{Tang2021b} $\approx -\sigma_{xz}^{y}$ [this work],
indicating that the crystalline structure of the samples used in \cite{Tang2021b} is LHC rather than RHC considered here.
Nevertheless, the $\sigma_{xy}^{z}$ spectrum in \cite{Tang2021b} near Weyl point W$_2$ 
differs significantly from our $\sigma_{xz}^{y}$ spectrum. These discrepancies could reflect
the different computational methods and structural parameters used in the previous \cite{Tang2021b} and present studies. 
We note that the independent nonzero SHC element $\sigma_{xz}^{y}$ was not 
studied in Ref. \cite{Tang2021b}. 

Finally, we notice that for the application of SHE in spintronics such as spin-orbit torque
switching-based nanodevices, the crucial quantity is the so-called spin Hall angle $\Theta_{sH}$
which characterizes the charge-to-spin conversion efficiency and is given by $\Theta_{sH} = (2e/\hbar)J^s/J^c = 2\sigma^s/\sigma^c$ where
$J^c$ and $\sigma^c$ are the longitudinal charge current density and conductivity, respectively
(see, e.g., Refs. \cite{Tung2013} and \cite{Wang2014}). Therefore, although the calculated
SHC values of the CoSi family are much smaller than that of 5$d$ transition metals such as Pt
(Table II), their spin Hall angles could be comparable to that of 5$d$ transition metals~\cite{Wang2014} 
because topological semimetals by nature have a much smaller conductivity compared with
5$d$ transition metals (Table II). For example, the spin Hall angle $\Theta_{sH}^z$ for
$\sigma_{xy}^{z}$ of RhSi and RhGe is about -7 \%, being comparable to that (10 \%) of Pt (Table II). 
Therefore, we believe that this interesting finding of large spin Hall angles of the CoSi family 
would spur further experiments on SHE in the members of the CoSi family other than CoSi~\cite{Tang2021b}.

\subsection{Spin Nernst effect}

The SNC ($\alpha_{ij}^{s}$; $s,i,j=x,y,z$) of a material is also a third-rank tensor, 
thus having 27 tensor elements altogether. 
As for the SHC, owing to the high symmetry of the cubic lattice for the CoSi family, it has only 
two independent nonzero tensor elements, namely, $\alpha_{xy}^{z}$ 
and $\alpha_{xz}^{y}$ \cite{Gallego2019} (see also Table I).
The calculated values of these nonzero SNC elements 
at $T = 300$ K are listed in Table II.
Remarkably, Table II indicates that the calculated value of SNC $\alpha_{xz}^{y}$ of CoGe is 
comparable or even larger than that ($\alpha_{xy}^{z}$) of platinum metal (Table II). 
Also, $\alpha_{xz}^{y}$ of CoGe is about 2.5 times larger than $\alpha_{xz}^{y}$  
of Dirac line-node semimetal ZrSiS \cite{Yen2020} (Table II).
This shows that the SNC values for the CoSi family are prominent and the CoSi family would be potential materials 
for spin caloritronics.

Table II indicates that the SNC of the CoSi family is strongly anisotropic, similar to their SHC.
For example, the $\alpha_{xy}^{z}$ of CoGe is about 21 times larger than $\alpha_{xz}^{y}$ (see Table II), 
i.e., the SNC of CoGe would be enhanced by a factor of about $21$ when the applied electric field 
and spin polarization directions are interchanged from ($y$, $z$) to ($z$, $y$). Here the indices in the parentheses 
indicate the directions of the applied electric field and spin polarization, respectively.
Also, the $\alpha_{xy}^{z}$ of RhGe is approximately $3.4$ times larger than $\alpha_{xz}^{y}$ (Table II). 
A sign change in the SNC is found for CoSi, CoGe and RhSi (Table II) when the spin polarization
direction is rotated from the $z$-axis to $y$-axis.

We also calculate the chemical potential ($\mu$) 
dependence of the SNC at $T = 300$ K for CoSi, CoGe, RhSi and RhGe 
as shown in Figs. \ref{fig:SHC-XY}(c), \ref{fig:SHC-XY}(f), \ref{fig:SHC-XY}(i), and \ref{fig:RhGe}(c), respectively.
As for the SHC, the SNC spectra of the CoSi family also depend strongly on $\mu$.
For example, $\alpha_{xy}^{z}$ of CoGe becomes $19$ times larger and also changes sign 
when the chemical potential is lowered to the spin-1 double Weyl fermion node at $R$ (i.e., W$_2$ at -0.15 eV)
[see Fig. \ref{fig:SHC-XY}(f), Table II  and Table S1 in the SM \cite{Prasad-SM}].
This suggests that the presence of unconventional chiral fermion nodes may considerably enhance the SNE. 
We note that this chemical potential lowering can be realized by hole doping of 0.27 e/f.u. 
Also, Fig. \ref{fig:SHC-XY}(c) indicates that the $\alpha_{xy}^{z}$  of CoSi 
can be increased from 0.42 to -1.35 ($\hbar$/e)(A/m K)  
when $\mu$ is shifted to node W$_2$ (-0.18 eV) [see also Table S1]. 
This can be achieved via hole doping of 0.25 e/f.u.   
Furthermore, the SNC ($\alpha_{xz}^{y}$) of RhGe is enhanced by a factor of about $1.5$ when $\mu$ drops to 
the level of the spin-3/2 fermion node W$_1$ at $\Gamma$ (see Table II and Table S1 in the SM \cite{Prasad-SM})
which can be achieved  via hole doping of merely 0.06 e/f.u.
Nevertheless, we also notice that this enhancement in SNC is absent in  the $\alpha_{xy}^{z}$ of CoSi, CoGe and RhSi 
at W$_1$ (see Fig. \ref{fig:SHC-XY}, Table II and also Table S1 in the SM). 
Also, Fig. \ref{fig:RhGe}(c) shows that there is no local maximum in the $\alpha_{xy}^{z}$ near W$_1$ in RhGe.  
Overall, the $\mu$ dependences of the $\alpha_{xy}^{z}$ and $\alpha_{xz}^{y}$ of CoSi, CoGe and RhSi
are rather similar (see Fig. \ref{fig:SHC-XY}). 
For CoSi, the $\alpha_{xy}^{z}$ decreases steadily as $\mu$ decreases and changes sign at -0.10 eV, 
and then reaches a negative local maximum of -1.37 ($\hbar$/e)(A/m K) at -0.19 eV [see Fig. \ref{fig:SHC-XY}(c)].
On the contrary, $\alpha_{xy}^{z}$ of RhGe increases rapidly as $\mu$ decreases, 
changes sign at -0.08 eV and then rises to a positive local maximum of 0.67 ($\hbar$/e)(A/m K) at -0.16 eV
[Fig. \ref{fig:RhGe}(c)].
These remarkable tunabilities in the SNC could be observed in 
the CoSi family by either chemical substitution or electrical gating. 

We note that Eq. (\ref{eq:3}) would be reduced to the Mott relation at the low $T$ limit,
\begin{equation}
\label{eq:4}
\begin{aligned}
\alpha_{x y}^{z}\left(E_{F}\right)=-\frac{\pi^{2}}{3} \frac{k_{B}^{2} T}{e} \sigma_{x y}^{z}\left(E_{F}\right)^{\prime},
\end{aligned}
\end{equation}
which simply means that the SNC is proportional to the energy derivative of the SHC at $E_F$.
This would allow us to understand the origins of the prominent features in the $\mu$-dependent 
SNC spectra. For example, the large negative peak in the $\alpha_{xz}^{y}$ spectrum of CoSi 
in the vicinity of node W$_1$ is due to the positive steep slope of $\sigma_{xz}^{y}$ 
near W$_1$ [see Figs. \ref{fig:SHC-XY}(b) and \ref{fig:SHC-XY}(c)]. 
This large SNC ($\alpha_{xz}^{y}$) of -1.07 ($\hbar$/e)(A/m K) at $T=300$ K is located at 0.02 eV,
being slightly above $E_F$, and thus could be achieved easily via electron doping of 0.01 e/f.u.
Similarly, in Figs. \ref{fig:SHC-XY}(b)-(c), $\alpha_{xy}^{z}$ of CoSi also has a peak value of 
-1.37 ($\hbar$/e)(A/m K) at $T=300$ K at spin-1 double Weyl fermion node W$_2$ ($\mu =$ -0.19 eV) at $R$
[Fig. \ref{fig:SHC-XY}(c)] and again this is due to the steep positive slope 
of $\sigma_{xy}^{z}$ near W$_2$. In order to reach this energy level, however, a larger hole doping 
of 0.27 e/f.u. would be required. In contrast, Fig. \ref{fig:SHC-XY}(f) shows that for CoGe, $\alpha_{xy}^{z}$ 
is negligibly small above $E_F$ because $\sigma_{xy}^{z}$ is very flat in this energy range 
[Fig. \ref{fig:SHC-XY}(e)].  
Unlike CoSi, CoGe and RhSi, both $\alpha_{xy}^{z}$ and $\alpha_{xz}^{y}$ of RhGe are negative
and  $\alpha_{xz}^{y}$ of RhGe has a broad plateau of about -0.18 ($\hbar$/e)(A/m K) at $T=300$ K
from -0.05 to 0.10 eV around $E_F$ [see Fig. \ref{fig:RhGe}(c)], where $\sigma_{xz}^{y}$ increases rather slowly with $\mu$ [see Fig. \ref{fig:RhGe}(b)].

\begin{figure}[htbp] \centering
\includegraphics[width=8.0cm]{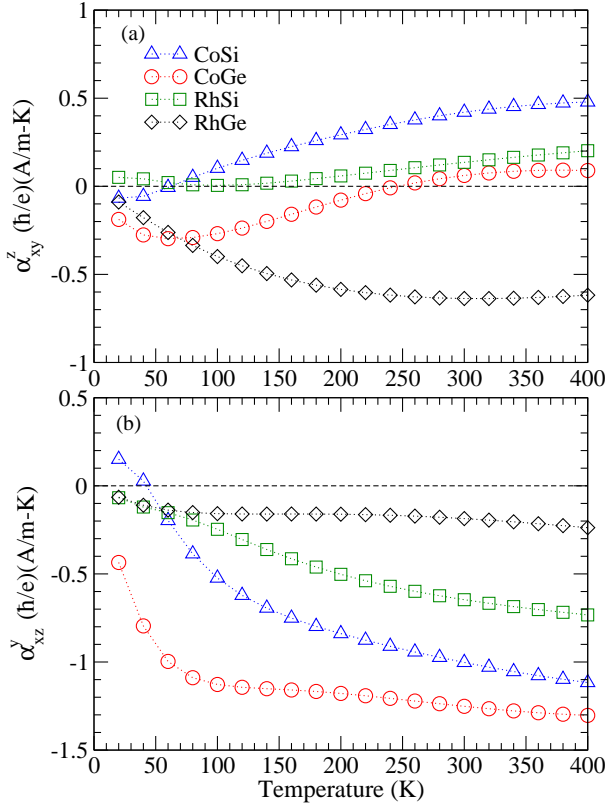}
\caption{Spin Nernst conductivity (a) $\alpha_{xy}^{z}$ and (b) $\alpha_{xz}^{y}$ of the CoSi family 
as a function of temperature $T$.}
\label{fig:T-scan}
\end{figure}
 
The temperature ($T$) dependence of the SNC ($\alpha_{xy}^{z}$ and $\alpha_{xz}^{y}$) 
of the CoSi family is also calculated, as shown in Fig. \ref{fig:T-scan}. 
In Figs. S11 and S12 in the SM \cite{Prasad-SM}, we also display the $T$ dependences of $\alpha_{xy}^{z}$, 
and $\alpha_{xz}^{y}$ when $\mu$ is displaced to nodes W$_1$ and W$_2$, respectively. 
Figure \ref{fig:T-scan}(a) shows that the $\alpha_{xy}^{z}$ spectra of CoSi and RhSi have 
a positive value for almost the entire considered $T$ range, which increases steadily with $T$ 
and reaches 0.48 and 0.20 ($\hbar$/e)(A/m K), respectively, at $T=400$ K.  
For CoGe, $\alpha_{xy}^{z}$ has a negative value of -0.30 ($\hbar$/e)(A/m K) at $T=60$ K. 
As $T$ increases, $\alpha_{xy}^{z}$ increases and changes sign at $T = 240$ K and then rises to a value 
of 0.09 ($\hbar$/e)(A/m K) at $T=400$ K [see Fig. \ref{fig:T-scan}(a)]. 
However, $\alpha_{xy}^{z}$ of RhGe has a negative value in the whole $T$ range considered here. 
Its magnitude increases monotonically and hits the negative maximum value of -0.64 ($\hbar$/e)(A/m K) 
at $T=320$ K. It then starts to decrease as $T$ further increases and finally reduces 
to -0.62 ($\hbar$/e)(A/m K) at $T=400$ K [see Fig. \ref{fig:T-scan}(a)].
In Fig. \ref{fig:T-scan}(b), unlike the $\alpha_{xy}^{z}$, $\alpha_{xz}^{y}$ has a negative value for 
all four investigated compounds of the CoSi family in almost the entire considered temperature range.
For CoSi, RhSi and CoGe, the magnitude of $\alpha_{xz}^{y}$ increases monotonically as $T$ increases from 50 K, 
and reaches the negative value of -1.12, -0.73 and -1.30 ($\hbar$/e)(A/m K), respectively, at $T=400$ K 
[see Fig. \ref{fig:T-scan}(b)]. $\alpha_{xz}^{y}$ of RhGe shows a robust behavior with $T$ 
and decreases slightly to -0.24 ($\hbar$/e)(A/m K) at $T=400$ K, as can be seen easily in Fig. \ref{fig:T-scan}(b).
Interestingly, for RhGe, $\alpha_{xy}^{z}$ changes sign with $T$ when $\mu$ is shifted to node W$_1$ 
[see Fig. \ref{fig:T-scan}(a) and Fig. S11(a)]. 
Also, for the CoSi family, the $\alpha_{xy}^{z}$ is negative whereas $\alpha_{xz}^{y}$ is positive 
for the entire considered $T$ range when $\mu$ is lowered to node W$_2$ (see Fig. S12 in the SM \cite{Prasad-SM}).

Finally, from the viewpoint of the application of SNE in spin caloritronics, the key quantity is 
the spin Nernst angle $\Theta_{sN}$ which measures the heat-to-spin conversion efficiency and 
is defined as $\Theta_{sN} = (2e/\hbar)J^s/J^h = 2\alpha^s/\alpha^L$ where
$J^h$ and $\alpha^L$ are the longitudinal heat current density and Nernst coefficient, respectively~\cite{Meyer2017}.
Here $\alpha^L = S_{xx}\sigma_{xx}$ where $S_{xx}$ is the Seebeck coefficient.
Using the measured $\sigma_{xx}$ and $S_{xx}$ of the CoSi family, we estimate the $\Theta_{sN}$
values using the calculated $\alpha^s$, as listed in Table II. Interestingly, we obtain the large
$\Theta_{sN}$ values of 15 \% and 12 \%, respectively, for RhSi and RhGe (Table II), being comparable to that (-20 \%) 
of Pt metal~\cite{Meyer2017}. So far no experimental nor theoretical studies on
the SNE in the CoSi family have been reported. We hope that this interesting finding of large spin Hall (Nernst) angles
in the CoSi family especially RhSi and RhGe would stimulate experiments of this kind in the near future.

\subsection{Spin Berry curvature analysis}

\begin{figure}[htbp] \centering 
\includegraphics[width=8.6cm]{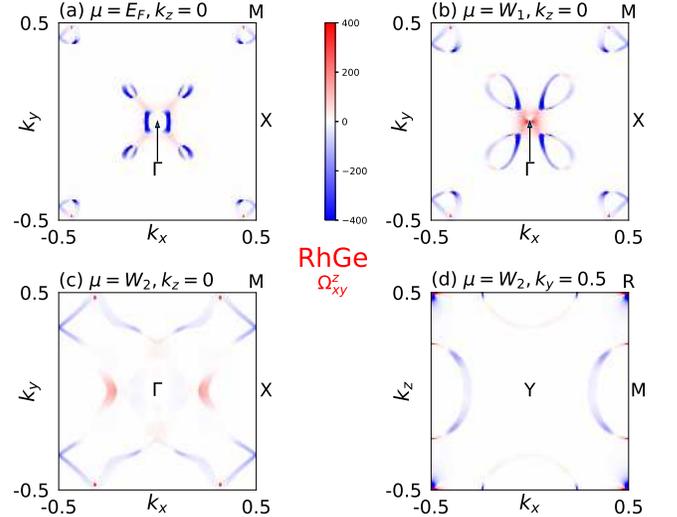}
\caption{${\bf k}$-resolved spin Berry curvature $\Omega_{xy}^{z}({\bf k})$ of RhGe 
on the $k_{x}$-$k_{y}$ plane with $k_{z}=0$ (a, b, c) and also on the $k_{x}$-$k_{z}$ plane with $k_{y}=0.5$ 
(d) in the Brillouin zone. The color bar is in units of \AA$^{2}$. In (a), chemical potential $\mu = E_F$; 
In (b), $\mu = W_1$ (spin-3/2 fermion node at $\Gamma$); In (c) and (d), $\mu = W_2$ 
(spin-1 double Weyl fermion node at $R$).} 
\label{fig:RhGe-k-slice-xyz}
\end{figure} 

\begin{figure}[htbp] \centering 
\includegraphics[width=8.6cm]{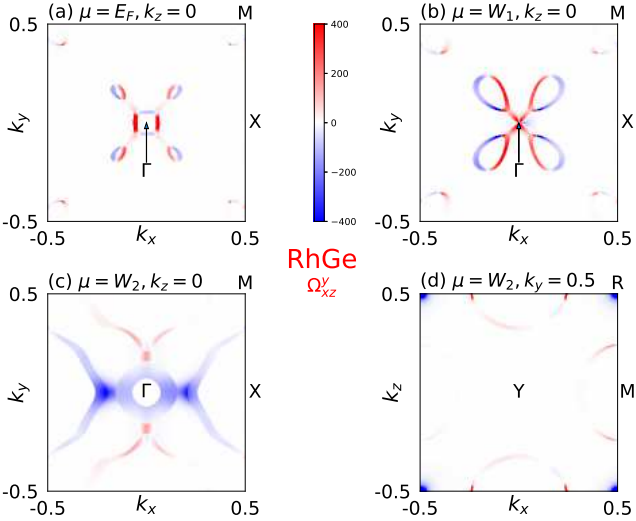}
\caption{${\bf k}$-resolved spin Berry curvature $\Omega_{xz}^{y}({\bf k})$ of RhGe 
on the $k_{x}$-$k_{y}$ plane with $k_{z}=0$ (a, b, c) and also on the $k_{x}$-$k_{z}$ plane with $k_{y}=0.5$ 
(d) in the Brillouin zone. The color bar is in units of \AA$^{2}$. In (a), chemical potential $\mu = E_F$; 
In (b), $\mu = W_1$ (spin-3/2 fermion node at $\Gamma$); In (c) and (d), $\mu = W_2$ 
(spin-1 double Weyl fermion node at $R$).}
\label{fig:RhGe-k-slice-xzy}
\end{figure}

We can see from Eq. (\ref{eq:1}) that the SHC is simply given by the summation of 
spin Berry curvature (SBC) of the occupied bands on all the $k$-points in the BZ. 
As a result, analysing the $k$-resolved SBC would help us understand 
the origins of the large SHC as well as SNC of the CoSi family. Table II indicates that 
RhGe has the largest SHC among the four considered compounds. 
Therefore, taking RhGe as an example, we display its calculated SBC $\Omega_{xy}^{z}$ 
in Figs. \ref{fig:RhGe}(d), \ref{fig:RhGe}(f) and \ref{fig:RhGe}(h) as well as $\Omega_{xz}^{y}$ 
in Figs. \ref{fig:RhGe}(e), \ref{fig:RhGe}(g) and \ref{fig:RhGe}(i) along the high-symmetry lines 
in the BZ for chemical potential $\mu$ = $E_F$, $\mu$ = W$_{1}$, and $\mu$ = W$_{2}$, respectively.
In order to get a clearer picture of the SBC distribution in the BZ,
we also show on the $k_x$-$k_y$ plane with $k_z=0$ the contour plots of $\Omega_{xy}^{z}$ in
Figs. \ref{fig:RhGe-k-slice-xyz}(a), \ref{fig:RhGe-k-slice-xyz}(b) and \ref{fig:RhGe-k-slice-xyz}(c)
and also of $\Omega_{xz}^{y}$ in Figs. \ref{fig:RhGe-k-slice-xzy}(a), \ref{fig:RhGe-k-slice-xzy}(b) 
and \ref{fig:RhGe-k-slice-xzy}(c) for  $\mu$ = $E_F$, $\mu$ = W$_{1}$, and $\mu$ = W$_{2}$, respectively.
Contour plots of $\Omega_{xy}^{z}$ and $\Omega_{xz}^{y}$ on the $k_x$-$k_z$ plane with $k_y=0.5$
for $\mu$ = W$_{2}$ are also presented in Figs. \ref{fig:RhGe-k-slice-xyz}(d) 
and \ref{fig:RhGe-k-slice-xzy}(d), respectively.

For $\mu$ = $E_F$, a sharp negative (positive) peak near the $\Gamma$ point on the $X$-$\Gamma$ line 
is found for $\Omega_{xy}^{z}$ ($\Omega_{xz}^{y}$) [see Figs. \ref{fig:RhGe}(d) and \ref{fig:RhGe}(e)]. 
Comparison of Fig. \ref{fig:RhGe}(a) with Fig. S1(d) 
indicates that the bands near $E_F$ around this $k$-point are slightly split when the SOC is included. 
As pointed before \cite{Guo2008}, when two degenerate bands become slightly gapped by the SOC, 
a pair of large peaks of SBC with opposite signs would occur in
the vicinity of this $k$-point. When both bands are occupied, the contributions from these two peaks to the SHC 
would cancel each other. However, when $E_F$ falls within the gap, only one peak of SBC 
would contribute to the SHC, thus resulting in a pronounced contribution to the SHC.
Therefore, we can see that the large negative (positive) peak of $\Omega_{xy}^{z}$ ($\Omega_{xz}^{y}$) 
makes a crucial contribution to the SHC, leading to the large negative $\sigma_{xy}^{z}$ (positive $\sigma_{xz}^{y}$) value
at $E_F$. These large negative $\Omega_{xy}^{z}$ and positive $\Omega_{xz}^{y}$ peaks near $\Gamma$ 
can be seen more clearly in Figs. \ref{fig:RhGe-k-slice-xyz}(a) and \ref{fig:RhGe-k-slice-xzy}(a), respectively. 
Figure \ref{fig:RhGe}(d) also shows that there are less prominent negative peaks of $\Omega_{xy}^{z}$ 
along the $M$-$\Gamma$ [see also Fig. \ref{fig:RhGe-k-slice-xyz}(a)] and $\Gamma$-$R$ lines, 
which should also contribute significantly to the large negative $\sigma_{xy}^{z}$ value.
Again, these large $\Omega_{xy}^{z}$ values originate from the small SOC-induced band splitting
near $E_F$ [see Fig. \ref{fig:RhGe}(a)]. Similarly, one can find rather pronounced positive $\Omega_{xz}^{y}$ peaks
along the $\Gamma$-$R$ line. 

When the chemical potential is lowered to the spin-3/2 fermion node ($\mu$ = W$_{1}$) at $\Gamma$, 
a large asymmetric oscillator-like curve occurs for $\Omega_{xy}^{z}$ near $\Gamma$ on the $X$-$\Gamma$-$R$ line
[see Fig. \ref{fig:RhGe}(f)], perhaps a unique feature of SBC near a chiral fermion node.
Nevertheless, the negative peak near $\Gamma$ on the $\Gamma$-$R$ side is gigantic, 
which overcomes the less pronounced positive peak on the $\Gamma$-$X$ side [Fig. \ref{fig:RhGe}(f)], 
thus resulting in the large negative $\sigma_{xy}^{z}$ value (Table II). 
Of course, other pronounced negative peaks in $\Omega_{xy}^{z}$ such as that along the $R$-$M$-$\Gamma$ line
may also contribute significantly to $\Omega_{xy}^{z}$.
Similarly, we can also see large oscillator-like features in $\Omega_{xz}^{y}$ near $\Gamma$ along the $X$-$\Gamma$-$R$ 
and $M$-$\Gamma$-$Y$ lines [see Fig. \ref{fig:RhGe}(g)]. However, clearly, the large positive peaks on
the $\Gamma$-$R$ and $\Gamma$-$M$ sides would overcome the smaller negative peaks of $\Omega_{xz}^{y}$ on
the $\Gamma$-$X$ and $\Gamma$-$Y$ sides, thus leading to the large positive $\sigma_{xz}^{y}$ value.
When the chemical potential is further lowered to the spin-1 double Weyl fermion node ($\mu$ = W$_{2}$) at $R$,
similar oscillatory features in both $\Omega_{xy}^{z}$ and $\Omega_{xz}^{y}$ appear
[see Figs. \ref{fig:RhGe}(h) and \ref{fig:RhGe}(i)]. 
The gigantic negative peaks of $\Omega_{xy}^{z}$ and $\Omega_{xz}^{y}$ on the $R$-$M$ side
[see Figs. \ref{fig:RhGe-k-slice-xyz}(d) and \ref{fig:RhGe-k-slice-xzy}(d)] 
would overwhelm the much smaller positive peaks
on the $R$-$\Gamma$ side, thereby resulting in the negative $\sigma_{xy}^{z}$ and $\sigma_{xz}^{y}$ values
[see Fig. \ref{fig:RhGe}(b)]. Of course, there is also a broad negative peak of $\Omega_{xz}^{y}$ 
on the $X$-$\Gamma$ line, being clearly due to the large SOC-split bands [see Fig. \ref{fig:RhGe}(a)],
and this negative peak shows up clearly as a small part of a complex negative feature
in the contour plot on the $k_x$-$k_y$ plane with $k_z =0$ [see Fig. \ref{fig:RhGe-k-slice-xzy}(c)].

\section{CONCLUSIONS}

In summary, we have systematically studied the band structure topology, SHE and SNE in the CoSi family
with the chiral cubic B20 structure 
by performing {\it ab initio} relativistic band structure calculations. 
First, all the four considered monosilicides (CoSi, CoGe, RhSi and RhGe) are found to
be nonmagnetic semimetals with unconventional chiral fermion nodes.
Second, unlike nonchiral cubic metals, these monosilicides have two independent
nonzero spin Hall (Nernst) conductivity tensor elements, namely,  $\sigma_{xy}^{z}$ and $\sigma_{xz}^{y}$
($\alpha_{xy}^{z}$ and $\alpha_{xz}^{y}$) instead of one element. 
Furthermore, the SHC ($\sigma_{xy}^{z}$ and $\sigma_{xz}^{y}$) and helicity of the chiral cubic structure are revealed to be correlated, thus suggesting SHE detection of structural helicity and also chiral fermion chirality. 
Third, the intrinsic SHE and SNE in some of the CoSi family are large. For example, the calculated SHC 
of RhGe is as large as -139 ($\hbar$/e)(S/cm). The calculated SNC of CoGe at room temperature is also
large, being -1.25 ($\hbar$/e)(A/m K). Because of their semimetallic nature with low electrical conductivity,
these topological semimetals may have large spin Hall and spin Nernst angles up to 7 \% and 15 \%, respectively, 
being comparable to that of platinum metal which has the largest SHC among transition metals. 
The SHC and SNC of these compounds can also be significantly enhanced  by raising or
lowering chemical potential to, e.g., spin-3/2 chiral fermion or spin-1 double Weyl node, 
via either chemical doping or electrical gating. 
Finally, an analysis of the calculated $k$-resolved spin Berry curvature unveils the mechanism 
underlying the largeness and tunability of SHE and SNE in these materials.
These interesting results thus show that transition metal monosilicides of the CoSi family not only
would provide a material platform for exploring novel spin-transports and exotic phenomena in
unconventional chiral fermion semimetals but also could be promising materials for developing better
spintronic and spin caloritronic devices.
We are sure that this work would stimulate new experiments on SHE and SNE in these fascinating topological semimetals.

\section*{ACKNOWLEDGMENTS}

The authors acknowledge the support from the Ministry of Science and Technology, National Center for Theoretical Sciences, 
and the Academia Sinica in Taiwan. 
The authors also thank the National Center for High-performance Computing (NCHC) in Taiwan for the computing time.

\end{document}